# Ferromagnetism and structural phase transition in rhombohedral PrIr$_3$

Karolina Gornicka[1,2], Michal J. Winiarski[2], Robert J. Cava[3], Michael A. McGuire[1], and Tomasz Klimczuk[2]

[1]Materials Science and Technology Division, Oak Ridge National Laboratory, Oak Ridge, Tennessee, United States

[2]Faculty of Applied Physics and Mathematics and Advanced Materials Centre, Gdansk University of Technology, ul. Narutowicza 11/12, 80-233 Gdańsk, Poland

[3]Department of Chemistry, Princeton University, Princeton, New Jersey 08540, United States

gornickaka@ornl.gov, tomasz.klimczuk@pg.edu.pl

**Abstract**

The synthesis, structural, magnetic, thermal and transport properties are reported for polycrystalline PrIr$_3$. At room temperature PrIr$_3$ displays the rhombohedral space group *R-3m* and a PuNi$_3$ – type structure. At around 70 K a phase transition to a monoclinic *C2/m* structure is observed and continued cooling reveals temperature independent behavior of the unit cell volume. Further, PrIr$_3$ undergoes a paramagnetic to ferromagnetic transition with T$_C$ = 7.5 K. The temperature dependent magnetic susceptibility follows the Curie Weiss law with a positive Curie-Weiss temperature, and an effective moment that is close to the theoretical effective moment for a free Pr$^{+3}$ ion. This introduces further complexity into the behavior of PuNi$_3$ - type materials and highlights the importance of temperature-dependent structural studies to complement physical property measurements in intermetallic compounds.

**Introduction**

The study of intermetallic compounds with 4$f$ electrons continues to intrigue researchers due to their diverse physical properties, such as superconductivity[1–6], valence instability[6–11], heavy fermion behavior[10–13], long-range magnetic order[14–23], and quantum criticality. Among these, the family of compounds characterized by the general formula $R_{2m+n}T_{4m+5n}$, where $R$ is a rare earth element and $T$ is a transition metal, stands out for its structural complexity and rich physical properties. The parameters $m$ and $n$ determine the number of $MgCu_2$-type and $CaCu_5$-type blocks, respectively, and provide a systematic framework for understanding the structural variations and their influence on the properties. The rare earth elements play a critical role in shaping the behavior of these compounds and the choice of transition metal introduces additional degrees of freedom that affect the bonding characteristics and electronic structure, thereby influencing various physical properties such as superconductivity, electrical conductivity, and thermal behavior.

A complex example is the $RT_3$ system ($R_3T_9$, $m = n = 1$), which is known to form in different crystal structures, depending on the choice of rare-earth and transition metal elements[1,18,22,24–29]. Among all these compounds, $R$Ir$_3$ stands out in terms of crystal structure and physical properties. It has been reported that for $R$ = La, Ce, and Nd, compounds crystallize in the rhombohedral PuNi$_3$ – type structure[5,6,9,22,30] with contrasting physical properties. LaIr$_3$ and CeIr$_3$ show superconductivity with $T_c$ = 2.5[5,30] and ~2.5-3.4 K[6,9], respectively, while NdIr$_3$ orders ferromagnetically below 10.6 K[22]. A recent report on another member, PrIr$_3$, suggested that this compound does not belong to the rhombohedral family and forms with co-existence of two polymorphic phases: AuBe$_5$- and AuCu$_3$-type[31]. The refined lattice parameters for cubic PrIr$_3$ reported there, as well as estimated Curie temperature, are very similar to those reported for the PrIr$_2$ phase in ref.[7,32]. Thus, further study of this system seems warranted.

In this work, we report the formation and properties of PrIr$_3$ in the PuNi$_3$ structure type at room temperature. We find a previously unreported structural transition from the rhombohedral $R$-3$m$ (#166) to the monoclinic $C2/m$ (#12) structure observed below 70 K, and a ferromagnetic ground state below a Curie temperature of 7.5 K. This structural transition introduces further complexity into the behavior of PrIr$_3$ and highlights the importance of low temperature-dependent structural studies in understanding the properties of intermetallic compounds.

**Experimental details**

A polycrystalline sample of PrIr$_3$ was synthesized by arc melting a stoichiometric mixture of high-purity elements, i.e., Pr (4N, Onyxmet) and Ir (3N, Mennica Metale, Poland), on a water-cooled Cu hearth under an argon atmosphere using zirconium as a getter. After the initial melt, the sample button was turned and remelted three times to insure reaction among the constituents. The weight loss upon melting was lower than 1%. The as-cast sample was subsequently wrapped in molybdenum foil and heat treated at 1350 °C for 36 h under high vacuum ($10^{-6}$ torr). After cooling down, the resulting material was ground in an agate mortar, pressed into a pellet, and then annealed under high vacuum at 1375 °C for 36 h.

Temperature-dependent powder X-ray diffraction (pXRD) experiments were performed in a vacuum using a PANalytical X'Pert Pro MPD diffractometer, with monochromatic Cu K$\alpha_1$ radiation. The temperature control was accomplished with an Oxford PheniX closed-cycle helium cryostat. The crystal structure was analyzed by Rietveld refinement using the Topas software. Temperature and field-dependent magnetization measurements were performed using a Quantum Design EverCool II Physical Property Measurement System (PPMS) with a vibrating sample magnetometer (VSM) function. The temperature dependencies of the zero-field cooled (ZFC) and field cooled (FC) dc magnetizations were measured in applied fields of 100 and 1000 Oe. Resistivity and heat capacity measurements were performed on a PPMS Evercool II. Specific-heat measurements were carried out in zero field using the two-τ time-relaxation method. The sample was cut to a suitable size and mounted with the Apiezon N grease onto the α-$Al_2O_3$ measurement platform to ensure good thermal contact. Resistivity measurement was performed on a polished fragment of a sample by means of the standard four-probe technique, with four 57-μm-diameter platinum wire leads attached to the flat sample surface using conductive silver epoxy (Epotek H20E).

Density Functional Theory (DFT) calculations of electronic structures of both variants of $PrIr_3$ were performed using the Projector Augmented Wave (PAW) method as implemented in the Quantum Espresso 7.2 package. PAW sets were taken from the PSLibrary database. The localized 4f orbitals of Pr were treated as core states. The Perdew-Burke-Ernzerhof Generalized Gradient Approximation (PBE GGA) was used for the exchange-correlation potential. Calculations were performed using the experimental crystal structures (*T* = 300 K for the rhombohedral variant and T = 20 K for the monoclinic one) and a 7x7x4 k-point grid.

**Results and discussion**

The primary structural characterization of $PrIr_3$ was carried out by X-ray diffraction at room temperature (see Fig.1(a)). The previous report on $PrIr_3$[31] found that it is dimorphic, with two different cubic structures i.e., an $AuBe_5$-type phase (*F-43m*, No.216, *a* = 7.922(8) Å) and an $AuCu_3$ – type phase (*Pm-3m*, No.221, *a* = 3.799 (8) Å). Mondal et al.[31] also synthesized $PrIr_3$ by first arc melting and then annealing for 7 days at the temperature of 900°C. In contrast, however, a preliminary report[33,34] suggested that this binary compound belongs to the rhombohedral branch of the homologous series described above, as do other analogues with the light lanthanide metals, i.e., LaIr$_3$[5], CeIr$_3$[9], or NdIr$_3$[22]. Dan *et al.*[16] proposed for NdIr$_3$ that the PuNi$_3$ – type structure can be obtained by extreme sintering: arc melting → 1$^{st}$ heating above 1300°C for 36h → grounding and pelletizing → 2$^{nd}$ heating above 1300°C for 36h and two polymorphic phases of NdIr$_3$ coexist when the sample is not exposed to extreme annealing (arc melting → heating at 900°C for 7 days)[16]. The Rietveld analysis performed on our data collected at 300 K showed that the structure of PrIr$_3$ is perfectly described by the rhombohedral space group *R-3m*, No.166, confirming the hypothesis that the PuNi$_3$ – type structure can be obtained by annealing above 1300°C (see Fig.1(a)). The refined lattice parameters obtained are $a_r$ = 5.3344(2) Å and $c_r$ = 26.268(3) Å. As expected, due to the lanthanide contraction effect, the resulting lattice constants are smaller than reported for LaIr$_3$[5] and larger than for the compound containing Nd[22]. Very weak, additional reflections (the strongest reflection appears at around 39.1°, (Fig. 1) were also detected, which were identified as small amounts (5.4%) of PrIr$_2$ (*Fd-3m*, No.227, *a* = 7.65 Å) impurity.

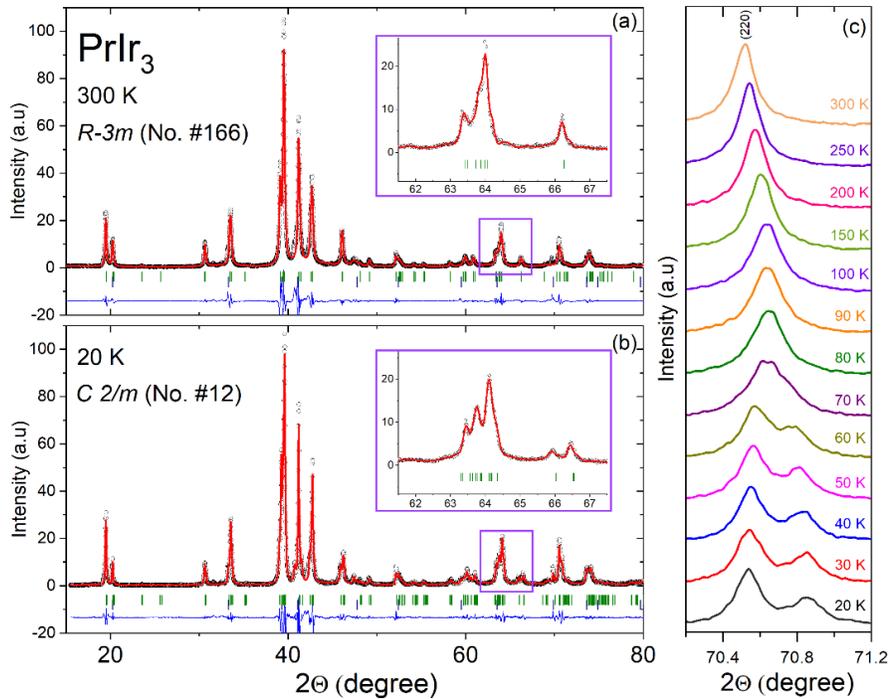

*Fig.1*. Powder X-ray diffraction patterns with the Rietveld profiles for PrIr$_3$ measured at 300 K (a) and 20 K (b). The green vertical marks show positions calculated for the Bragg reflections of the main phase. The second (black) vertical bars show positions for impurity phase PrIr$_2$. The lower trace is a plot of the difference between calculated and observed intensities. (c) Temperature evolution of the (220) reflection (*R-3m*) into the two reflections (*C2/m*).

The unit cell of the rhombohedral PuNi$_3$ – type PrIr$_3$ structure can be viewed as stacking of three types of layers, as shown in Fig.2(a): (breathing) kagome formed by Ir1, honeycomb Ir2 with triangular Pr2, and triangular Ir3 capped on both sides with two triangular Pr1 nets. Note, however, that the interlayer Ir-Ir distances are significantly shorter than the intralayer ones (e.g. Ir2-Ir2 distance is about 3.1 Å, while Ir2-Ir1 and Ir3-Ir2 - 2.7 Å), so the crystal structure cannot be considered quasi-two-dimensional.

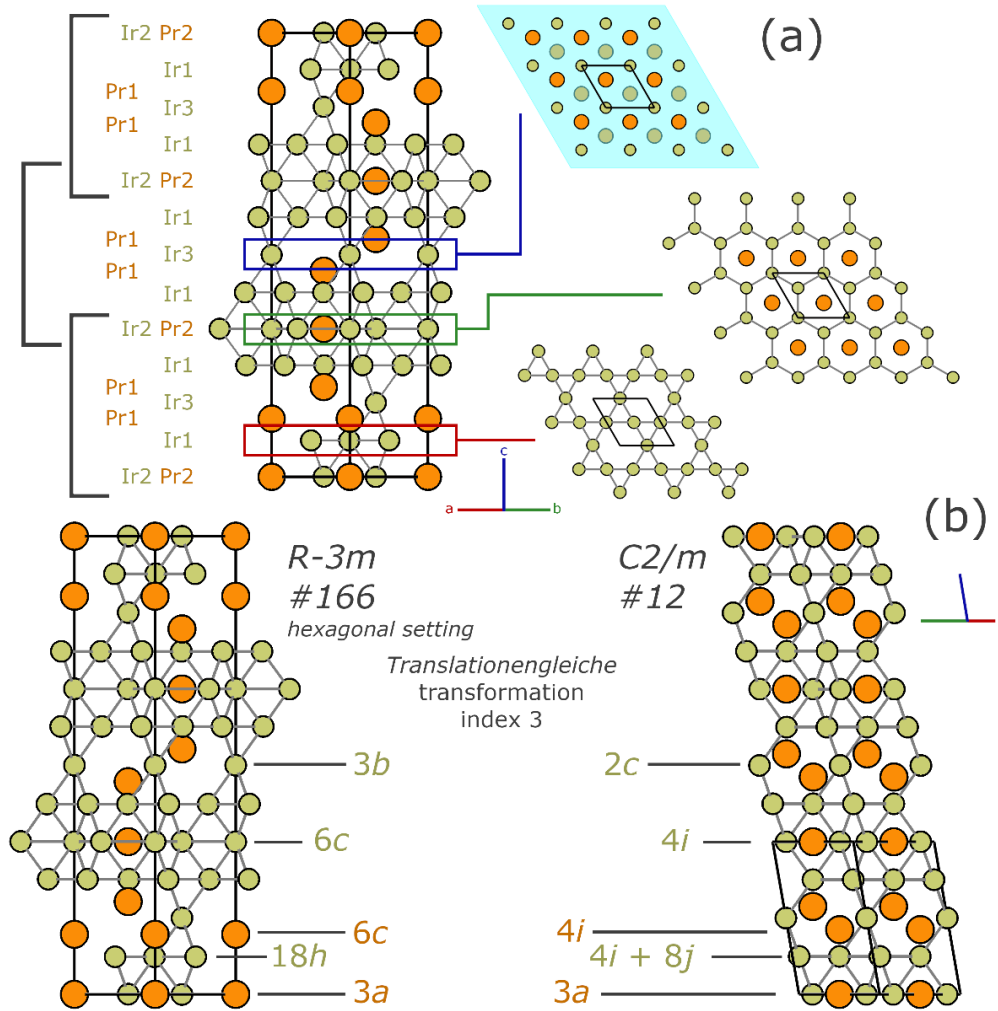

*Fig.2.* Panel (a) shows the high-temperature rhombohedral variant of the PrIr$_3$ structure decomposed to alternating layers: (breathing) kagome formed by Ir1 (Wyckoff position 18h; red rectangle), honeycomb Ir2 with triangular Pr2 (green rectangle), and triangular Ir3 capped with two triangular Pr1 nets (blue rectangle). Light blue rhombus is drawn to highlight that the Ir3 triangular net is capped by Pr1 atoms on both sides. Panel (b) shows the relationship between the rhombohedral *R-3m* (hexagonal setting) high-temperature phase cell and the distorted monoclinic (*C2/m*) low-temperature variant.

In rhombohedral NdIr$_3$, an anomaly in heat capacity and resistivity was observed at about 70 K[22]. A similar anomaly in physical property measurements is also observed in PrIr$_3$. To understand the nature of this transition, we performed temperature dependent measurements of the X-ray diffraction profiles to track the evolution of the crystal structure as a function of temperature. We identify several Bragg reflections that split or broaden with decreasing temperature (see Fig.1(b)). The increased number of Bragg reflections in the diffraction pattern indicates a lowering of the symmetry, and the nature of the peak splitting points to a loss of the trigonal symmetry. From a group/subgroup relationship argument (see Bärnighausen tree for *R-3m* from the Bilbao Crystallographic Server[35]), the most probable final space group for such a structural transition would be the monoclinic *C2/m* (No.12). The *translationengleiche* transformation from the rhombohedral *R-3m* to monoclinic *C2/m* space group removes the improper three-fold symmetry perpendicular to the stacked layers, leaving behind only the associated mirror plane *m*. The new two-fold axis is aligned along the monoclinic *b* axis, which is parallel to the *b* axis of the high-temperature unit cell.

| PrIr$_3$ at 300 K | | | | |
| --- | --- | --- | --- | --- |
| powder X-ray diffraction (Cu K$_\alpha$) | | | | |
| Unit cell parameter (Å) | | | | |
| $a_r$ | | | 5.3344(2) Å | |
| $c_r$ | | | 26.268(3) Å | |
| Atom | x | y | z | $B_{iso}$ (Å$^2$) |
| Pr1 | 0 | 0 | 0.1414(3) | 1.04(2) |
| Pr2 | 0 | 0 | 0 | 0.90(3) |
| Ir1 | 0.5002(4) | 0.4998(2) | 0.0829(2) | 0.89(5) |
| Ir2 | 0 | 0 | 1/3 | 0.60(8) |
| Ir3 | 0 | 0 | 1/2 | 0.76(4) |

*Table 1.* Refined structural parameters for PrIr$_3$ at 300 K. Space group *R-3m* (166). $B_{iso}$ is the thermal displacement parameter in square angstroms. Background-corrected Rietveld refinement reliability factors: profile residual $R_p$ = 18.8%, weighted profile residual $R_{WP}$ = 17.6%, expected residual $R_{exp}$ = 7.3%, goodness of fit $\chi^2$ =3.4.

A pXRD pattern collected at *T* = 20 K is shown together with the Rietveld refinement in Fig. 1(b). A list of refined structural parameters from powder XRD data collected at 300 and 20 K is presented in Tables 1 and 2. The lattice parameters obtained in the whole temperature range are provided in the Supplementary Material (SM).

| PrIr$_3$ at 20 K | | | | |
|---|---|---|---|---|
| powder X-ray diffraction (Cu K$_\alpha$) | | | | |
| Unit cell parameter (Å) | | | | |
| $a_r$ | | | 9.2477(9) Å | |
| $b_r$ | | | 5.3083(5) Å | |
| $c_r$ | | | 9.2566(8) Å | |
| $\beta$ | | | 109.04(6) ° | |
| Atom | x | y | z | B$_{iso}$ (Å$^2$) |
| Pr1 | 0.3594(3) | 0 | 0.5742(2) | 0.93(7) |
| Pr2 | 0 | 1/2 | 0 | 0.85(3) |
| Ir1 | 0.0835(4) | 0 | 0.2475(6) | 0.67(2) |
| Ir2 | 0.3323(5) | 0.2482(2) | 0.2477(8) | 0.81(5) |
| Ir3 | 0.1668(3) | 0 | 0 | 0.76(4) |
| Ir4 | 0 | 0 | 1/2 | 0.80(3) |

*Table 2.* Refined structural parameters for PrIr$_3$ at 20 K. Space group *C2/m* (12). B$_{iso}$ is the thermal displacement parameter in square angstroms. Background-corrected Rietveld refinement reliability factors: profile residual R$_p$ = 19.3%, weighted profile residual R$_{WP}$ = 18.1%, expected residual R$_{exp}$ = 8.4%, goodness of fit $\chi^2$ =3.8.

An example of the temperature intensity evolution of the Bragg reflection at 2θ = 70.60° is shown in Fig.1(c). Note, that the curves are offset to avoid overlapping. The diffraction reflection, which is indexed to be (220) for the rhombohedral phase, splits into two below 70 K for the monoclinic crystal structure. This can be illustrated by observing the evolution of the unit cell parameters with temperature as shown in Fig.4. For better comparability, in Fig. 4 we present the unit cell dimensions using the relation between the rhombohedral and monoclinic crystal structures: $a_r\sqrt{3} = a_m$ and $c_m \sin\beta = \frac{c_r}{3}$. This relationship is presented in Fig.3. We use the subscripts *r* and *m*, which stand for the crystal lattice constants for the high temperature (rhombohedral) and low temperature (monoclinic) form, respectively.

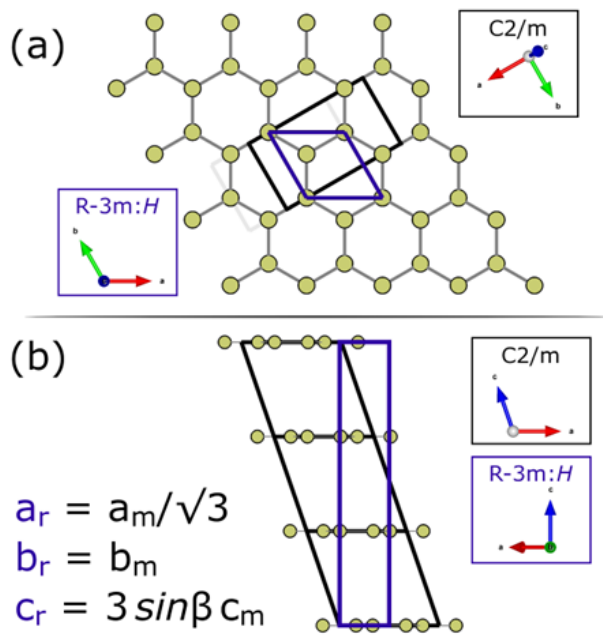

*Fig.3.* Relationship between rhombohedral (in hexagonal setting; purple) and monoclinic (black) cells and cell parameters. For clarity, only the honeycomb Ir2 sublattice is shown.

As can be seen in Fig. 4 (a), the variation of the lattice parameters with decreasing temperature is nearly linear in the temperature range of 100 – 300 K, in accordance with the normal thermal contraction of the unit cell. The smooth change in the c lattice constant through the transition temperature suggests that the transition is second order. This is further supported by the heat capacity measurements, which show no latent heat associated with the transition. The relative changes in the lattice parameters, $\Delta a/a$, $\Delta b/b$, $\Delta c/c$, near the structural transition are ~ 0.3%, -0.2%, and -0.01%, respectively. The $a_m$ axis exhibits a negative thermal expansion (note that am increases with decreasing temperature), suggesting that the structure is more disordered at low temperatures. We observe similar behavior for NdIr$_3$ (see Fig. 4(b)).

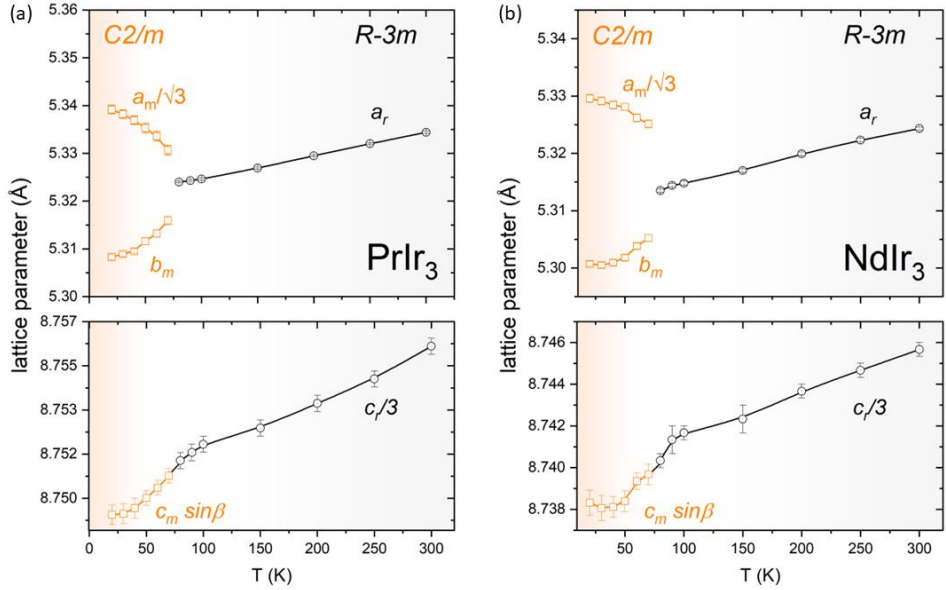

*Fig.4.* Lattice parameters *a,b* (upper panel) and *c* (lower panel) as a function of temperature. Note that the unit cell dimensions are presented according to the following relation: $a_r\sqrt{3} = a_m$ and $c_m \sin\beta = \frac{c_r}{3}$.

Figure 5 presents temperature dependence of the unit cell volume V(T) for PrIr$_3$ from 20 K to room temperature. Above 70 K, in the trigonal form, the volume of PrIr$_3$ behaves normally, i.e. V(T) increases with increasing temperature. However, in the monoclinic phase (below 70 K) the volume is nearly temperature independent.

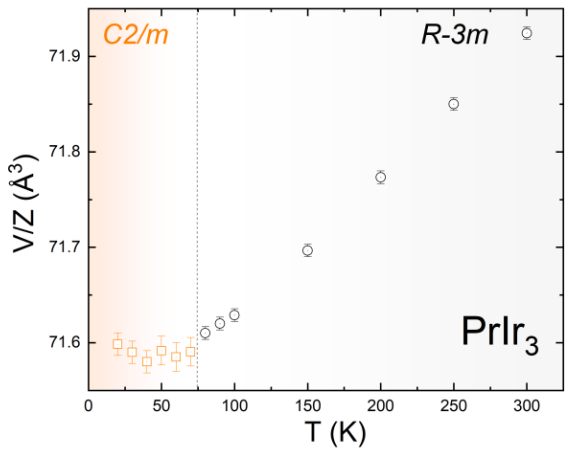

*Fig.5.* Temperature dependence of the unit cell volume divided by Z, where Z = 9 for the *R-3m* (open circles) and Z = 6 for the *C2/m* (open squares) crystal structure. The vertical dotted line marks the temperature of the structural phase transition.

Magnetization measurements provide details on the type of magnetic ordering in PrIr$_3$. The temperature dependence of magnetic susceptibility ($\chi$ = M/H) and inverse magnetic susceptibility (1/$\chi$) in an applied magnetic field of H = 1 kOe., from 2 to 300 K, are shown in Fig.6. The data were collected under zero-field cooled (ZFC) conditions, where the sample was cooled in the zero applied field and the magnetization (M) was measured while warming. No anomaly in the magnetic susceptibility is seen at the structural phase transition, which occurs in the paramagnetic phase, indicating that the magnetic moments and interactions change little at the transition. As shown in the main panel of Fig.6, there is a sharp increase in $\chi$ as the temperature is lowered below 8 K, suggesting a transition into a ferromagnetic state. The plot of inverse $\chi$(T) (see the inset in Fig.6) is found to be linear above the magnetic transition. In the paramagnetic state the data were analyzed using the modified Curie-Weiss law: $\chi(T) = \chi_0 + \frac{C}{T- \theta_P}$, where $\theta_P$ is known as the paramagnetic Curie-Weiss temperature, and C is referred to as the Curie constant. The temperature independent susceptibility term $\chi_0$ = -3.23(3) × 10$^{-4}$ emu mol$^{-1}$ and comes from both the sample and a sample holder. The best fit was obtained in the temperature range 150 - 250 K, and gave C = 1.279 (2) emu K mol$^{-1}$ and $\theta_P$ = 7.1(3) K. The effective magnetic moment per Pr ion can be determined using the relation $\mu_{eff} = \left(\frac{3Ck_B}{N_A\mu_B^2}\right)^{1/2} \cong \sqrt{8C}\mu_B$, where $k_B$ is the Boltzmann constant, $\mu_B$ is the Bohr magneton, and $N_A$ is Avogadro's number. The resulting effective magnetic moment for PrIr$_3$ $\mu_{eff}$ = 3.20(4) $\mu_B$/Pr is slightly smaller than the theoretical effective magnetic moment for a free Pr$^{+3}$ ion ($g_J$ [J(J+1)]$^{1/2}$ = 3.58$\mu_B$, where J and $g_J$ are the total angular momentum and its g-tensor, respectively)[36]. The magnitude of the paramagnetic Curie-Weiss temperature is related to the strength of the molecular field, which can be taken as an approximate indicator of the strength of the magnetic correlations between ions. When the molecular field aligns with the external field, the ferromagnetic interactions are dominant and a positive value of $\theta_P$ is observed. This is the case for PrIr$_3$, for which of $\theta_P = 5.3\ K$ has been obtained from the fit.

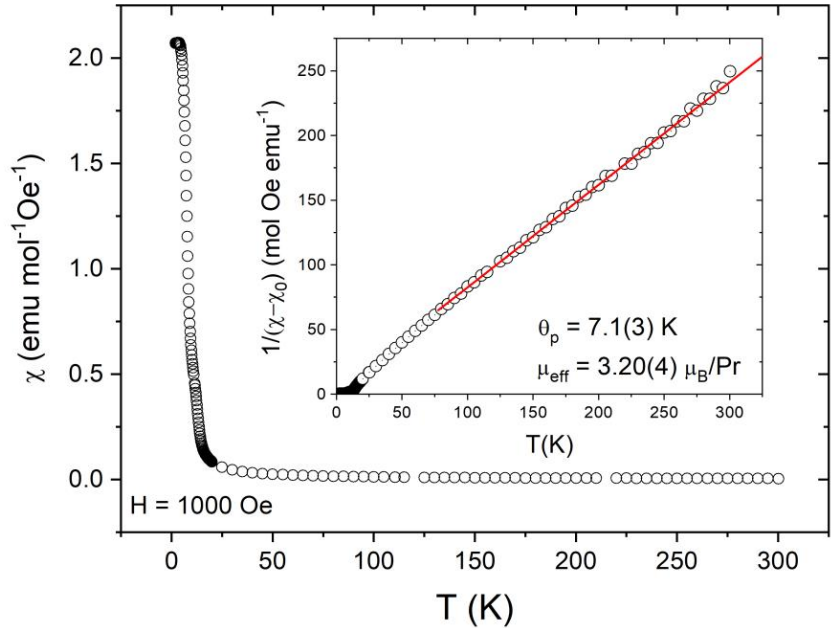

*Fig.6*. Temperature dependence of the ZFC magnetic susceptibility $\chi$ (T) in a field of H = 1 kOe. The inset shows the Curie-Weiss fit to the inverse magnetic susceptibility of PrIr$_3$ in 1 kOe, zero-field cooled state.

A comparison of zero-field-cooled (ZFC) and field-cooled (FC) magnetic susceptibility, under a field of H = 100 Oe and 1 kOe, for PrIr$_3$ is shown in Fig.7(a) and (b). At the lowest applied magnetic field (100 Oe), a sudden increase in $\chi$ (*T*) below 8 K is observed, which is indicative of a ferromagnetic ground state. As can be seen, the magnetic susceptibility is strongly dependent on the magnetic history of the sample, i.e., the ZFC curve decreases below a certain temperature, while the FC curve rises to near saturation at 2 K. The observed bifurcation in ferromagnets, in general, is related to the coercivity and the domain-wall pinning effect[36]. In a ferromagnetic system where the domain walls are particularly thin or narrow, if the applied magnetic field is less than the coercive field, the domain walls are pinned or frozen during cooling under ZFC conditions, resulting in a low net magnetization. On the contrary, as the temperature is lowered in the FC mode, the thermal energy facilitates the mobility of domain walls near the Curie temperature. As a result, the domains are aligned by a weak field, leading to enhanced magnetization. The difference between the ZFC and FC curves disappeared when the magnetic field was increased to 1 kOe (see panel (b) in Fig.7). A careful look at the curves reveals that there is an additional magnetic anomaly (a broad hump in the magnetic susceptibility centered near 11 K), due to the presence of a small amount of the impurity phase PrIr$_2$, which is a ferromagnet with the Curie temperature $T_C$ = 11.2(5) K[37].

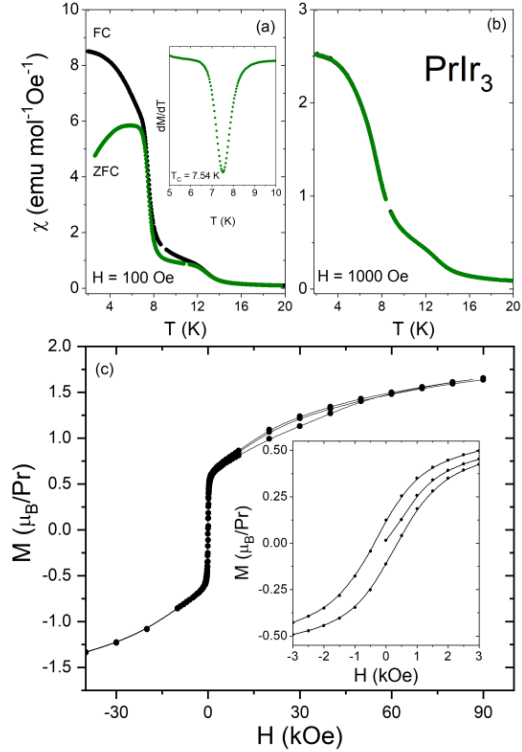

*Fig.7*. Temperature dependence of the ZFC and FC magnetic susceptibility $\chi$ (T) in a field of 100 Oe (a) and 1000 Oe (b). Inset in panel (a) shows the first order temperature derivative of the magnetization, identifying the Curie temperature of PrIr$_3$. (c) Magnetization isotherms measured at 2 K, in fields up to 90 kOe.

The Curie temperature ($T_C$) was determined from the first order temperature derivative ($dM/dT$) of the magnetization measured at 100 Oe applied field under ZFC conditions. For the ferromagnetic compounds, the minimum in the $dM/dT$ curve, measured at the low field, is often used to define the transition temperature ($T_C$) from the paramagnetic (PM) to the ferromagnetic (FM) state. As shown in the inset of Fig.7(a), $dM/dT$ for PrIr$_3$ reveals a negative peak centered at $T_C$ = 7.5 K, which is lower than the magnetic transition temperature reported for the isostructural NdIr$_3$ ($T_C$ = 10.6 K)[22]. The difference is most likely due to the lower value of the deGennes (dG) factor for the Pr ion as compared to the Nd ion.

To further evaluate the magnetic behavior, the field dependence of the isothermal magnetization M(*H*) was measured at *T* = 2 K (Fig.7(c)). Clearly, the M(*H*) dependence for PrIr$_3$ is characteristic of a ferromagnetic compound. The rather narrow hysteresis loop (H$_{coercive}$ = 370 Oe) may imply that PrIr$_3$ is a soft ferromagnet with a remnant magnetization ($\mu_R$ = 0.12$\mu_B$/Pr) being less than 10% of the magnetization at 9 T (see inset in Fig.7(d)). At *H* = 90 kOe the magnetization reaches about 1.65$\mu_B$, which is much smaller than the expected saturated value for Pr$^{+3}$ ($\mu_S = g_J J$ = 3.20$\mu_B$, where *J* is the total angular momentum and $g_J$ is the Landé g factor[36]). This saturation moment reduction may be caused by the crystal electric field (CEF) interactions and/or magnetocrystalline anisotropy generated by the anisotropic $Pr^{+3}$ ion, as is commonly observed in polycrystalline samples[38]. A lower than expected value of the saturation moment has also been found in $NdIr_3$[22], as well as in other Pr-based compounds[15,39].

In order to analyze the magnetic properties of $PrIr_3$ in detail, heat capacity measurements were carried out from $T$ = 300 to 2 K in order to identify phase transitions. The main panel of Fig.8(a) depicts the temperature dependence of the zero-field specific heat $C_p$ in the whole temperature range. At high temperatures, the specific heat is close to the classical high-$T$ Dulong-Petit limit $3nR \approx 100$ J mol$^{-1}$K$^{-1}$, where $n$ = 4 is the number of atoms per formula unit and $R$ = 8.314 J mol$^{-1}$K$^{-1}$ is the gas constant. A lambda anomaly at $T_C$ = 7.60 K is observed, as shown in more detail in Fig.7(b), confirming bulk character of the ferromagnetic transition.

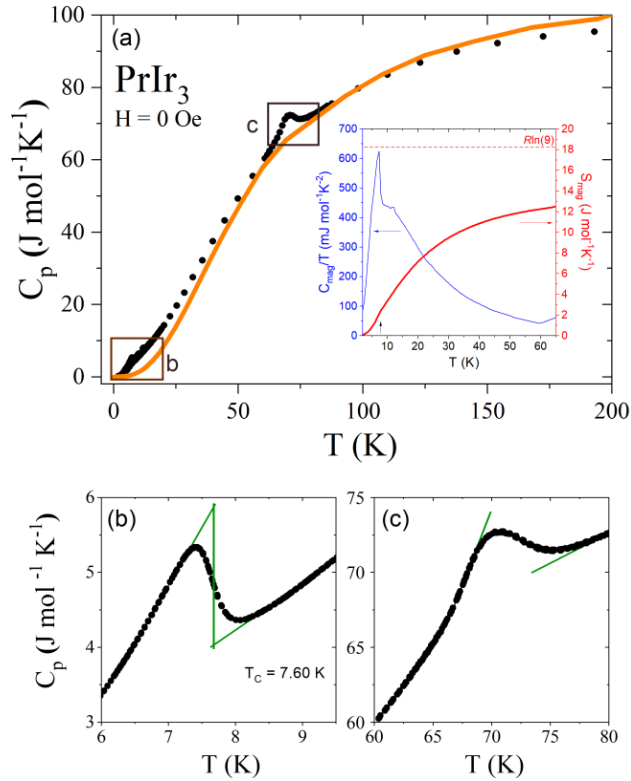

Fig.8 (a) Specific heat of $PrIr_3$ (black dots) and $LaIr_3$ (orange line) as a function of temperature. The inset shows the magnetic contribution ($C_{mag}/T$) and the magnetic entropy $S_{mag}$ as a function of $T$. (b) Expanded view of the heat capacity at the high-temperature structural transition. (c) The low-$T$ specific heat data in the vicinity of the ferromagnetic transition.

In addition to the specific heat jump at low temperature, a broad peak around 70 K (see panel (c) in Fig.8) is also noted. A similar feature was previously reported for isostructural $NdIr_3$[22]. The high-temperature anomaly observed in the heat capacity aligns well with the diffraction data and hence we conclude that it is related to a structural transition from a rhombohedral to a monoclinic crystal structure. The broad shape of the observed anomaly suggests a second-order transition. However, a broad peak observed in the specific heat may also be associated with a weak first-order transition. Such transitions are characterized by subtle discontinuities in the thermodynamic properties and can be manifested as a broad or rounded feature in the experimental data. This is often due to the system approaching a critical point where the transition exhibits features that are not as sharp as those of more pronounced first order transitions. In addition, a small amount of impurities can affect the shape of the peak observed in the specific heat data. To definitively rule out the possibility of a second-order phase transition, additional experiments, such as dilatometry, are required.

The magnetic contribution ($C_{mag}$) to $C_p$ for PrIr$_3$ was calculated by subtracting the $C_p$ of the nonmagnetic reference LaIr$_3$ (see the main panel in Fig.8(a)). The blue line in the inset of Fig.8(a) represents the calculated magnetic heat capacity ($C_{mag}/T$) that was used to get the magnetic entropy by using the expression $S_{mag} = \int \frac{C_{mag}}{T} dT$. The temperature dependent magnetic entropy for PrIr$_3$ is presented as a solid red line in the inset of Fig.8(a). It can be seen that the entropy increases continuously and saturates at higher temperatures to about 78% of the expected value for a $J = 4$ (Pr$^{+3}$), which is $R\ln(9)$ (not shown here). The discrepancy between the observed entropy value and the expected value for Pr$^{+3}$ can be attributed to several factors. Crystal electric field (CEF) effects should be considered as they can affect the energy levels and thermodynamic properties of Pr-containing compounds, potentially contributing to deviations from the expected entropy value[15,39–41]. Another factor is that the phonon contribution may be different below 70 K for the two compounds because they have different structures. Note that no anomaly is seen around 70 K for LaIr$_3$.

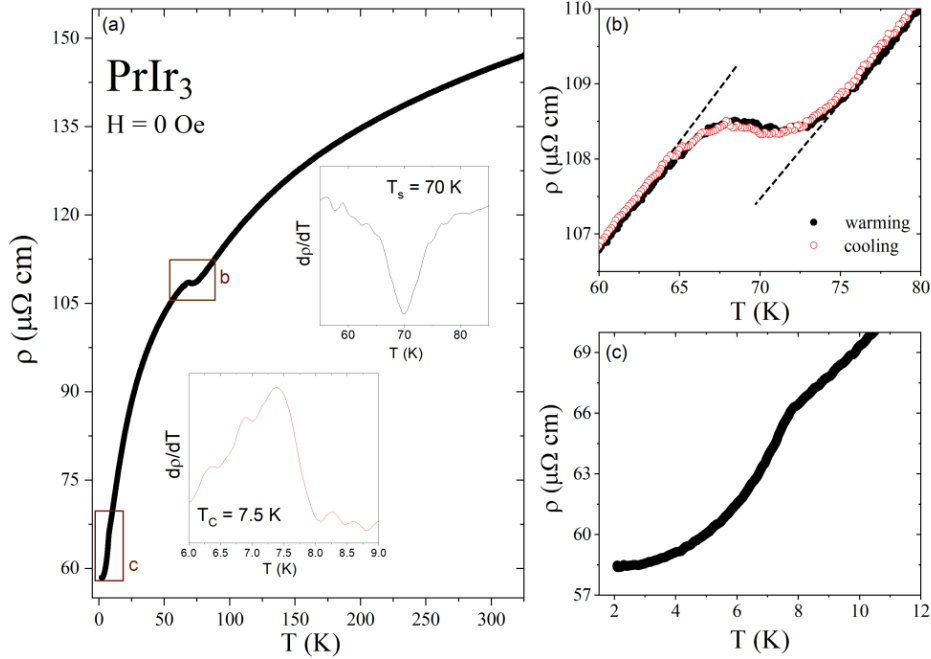

*Fig*.9 (a) Zero-field resistivity of PrIr$_3$ measured in the temperature range 2 - 300 K. The upper and lower insets show the first derivative vs. temperature in the vicinity of the structural and the magnetic transition, respectively. (b) Expanded view of the high-temperature anomaly. Data were collected during cooling and heating. (c) Temperature-dependent electrical resistivity *ρ(T)* in the vicinity of the magnetic transition.

To explore the electronic behavior, the temperature dependence of the electrical resistivity *ρ(T)* was measured from room temperature to 2 K in zero magnetic field (panel (a) in Fig.9). The resistivity shows metallic behavior with a slightly curvilinear trend. The *ρ(T)* deviates from linearity at high temperatures, with the temperature dependence progressively weakening. This behavior, commonly observed in alloys and intermetallic compounds, can be explained within the Mott-Ioffe-Regel limit[42]. This occurs when the mean free path (or average distance between collisions) is equal to the interatomic spacing, leading to the saturation of *ρ(T)* at high temperatures. The residual resistivity ratio (RRR) for PrIr$_3$ is ∼2.1, which is small compared to typical metals. Such a low value of RRR may appear due to the polycrystalline form of the sample (scattering on the grain boundaries) or by possible internal disorder at a level that is below the detection limit for our X-ray-diffraction technique. In addition, structural domains may form during the transition from the rhombohedral to the monoclinic structure. This could be related to the increase in resistivity on cooling through the transition (along with effects from the change in electronic structure) and could also contribute to the low RRR. The resistivity shows two features at 70 and 7.5 K (see insets in Fig.9(a), where the first derivative of the *ρ* vs. *T* is plotted). The anomaly at higher temperature is consistent with the broad peak observed in the specific-heat data. Panel (b) in Fig.9. shows the expanded view of the structural transition anomaly. It is worth noting that a detailed analysis of the heating-cooling *ρ(T)* curve does not reveal any discernible hysteresis loop, suggesting a second-order structural transition in PrIr$_3$ consistent with the group sub-group relationship noted above. The low-temperature anomaly is associated with PM-FM magnetic transition where resistivity drops rapidly due to reduction of spin scattering as the system orders magnetically (Fig.9(c)). The Curie temperature determined as the maximum of $d\rho/dT$ is equal to 7.5 K under zero magnetic field in very good agreement with the value obtained from the magnetic susceptibility and heat-capacity measurements.

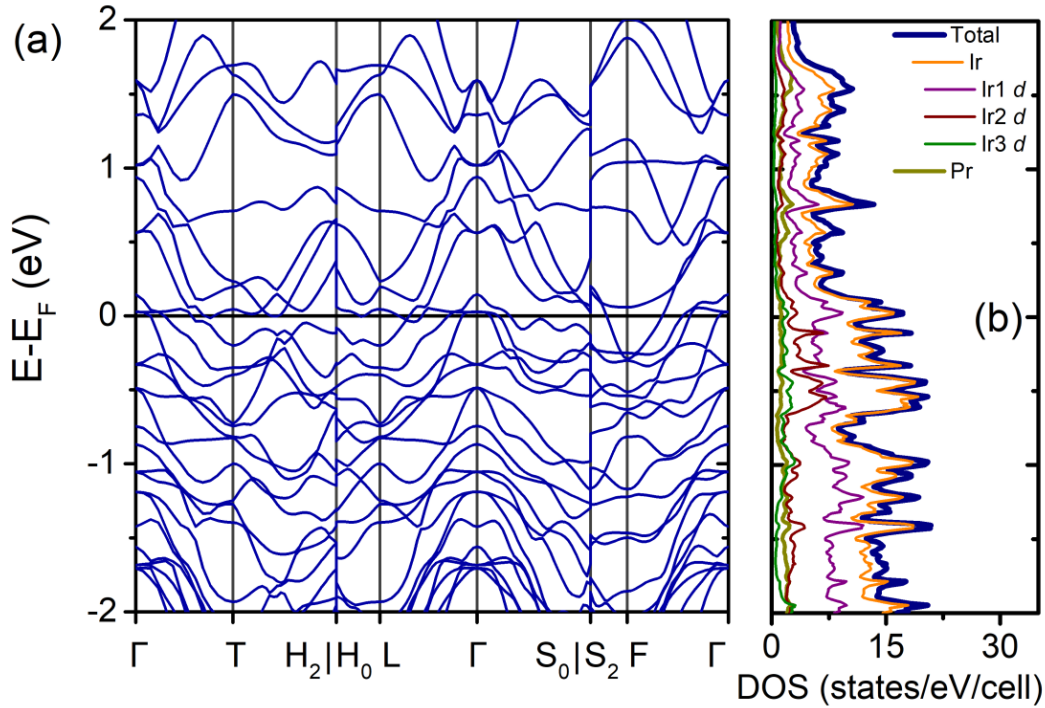

*Fig*.10 Band structure (a) and density of states (b) of the rhombohedral high-temperature (*T* = 300 K) variant of PrIr$_3$.

To better interpret the properties of PrIr$_3$ the electronic band structure and the density of electronic states (DOS) were calculated for both the high and low-temperature variants of the crystal structure. Since the calculations do not converge when the Pr 4*f* electrons are treated as valence electrons[43], we have treated the 4*f* electrons on Pr as core electrons. Electronic structure calculations (see Fig. 10 and 11) show that in both variants the dominating contribution to the density of states around the Fermi level DOS ($E_F$) come from the 5*d* states of the Ir atoms. Several weakly dispersive bands around $E_F$ result in a peak in thef DOS. By comparing the densities of states of these two systems, plotted in Figs. 10(b) and 11(b), one can see that the overall shape of the DOS function of PrIr$_3$ is similar.

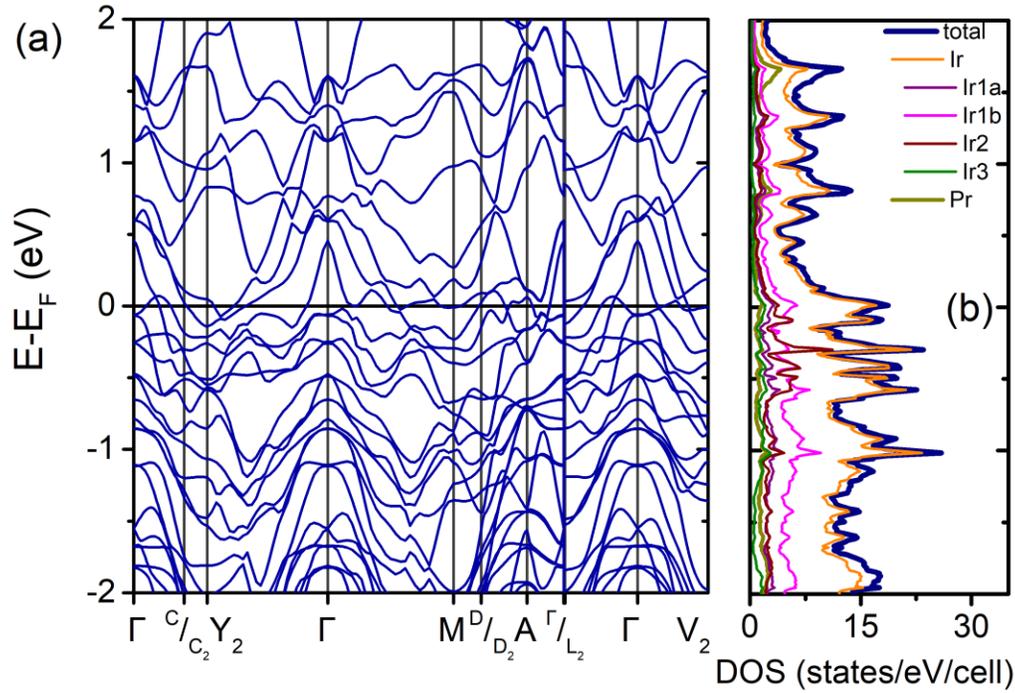

*Fig*.11 Band structure (a) and density of states (b) of the monoclinic low-temperature ($T$ = 20 K) variants of PrIr$_3$.

**Conclusions**

Our study provides comprehensive insights into the properties of the binary intermetallic compound PrIr$_3$, shedding light on its structural and magnetic behavior. We have shown that PrIr$_3$, with refined lattice parameters $a_r$ = 5.3344(2) Å and $c_r$ = 26.268(3) Å, belongs to the $R_{2m+n}T_{4m+5n}$ family of rhombohedral structures. As expected for the lanthanide series of compounds, the lattice parameters and unit cell volume are between those reported for NdIr$_3$[22] and LaIr$_3$[5].

Detailed physical property studies show presence of two-phase transitions for PrIr$_3$. The low temperature transition has been revealed by all three techniques we used, i.e. magnetic susceptibility, heat capacity and resistivity. The field dependent magnetization at 2 K shows a character typical for the ferromagnetic compounds with rather low coercive field (370 Oe). The bulk character of the PM-FM transition is confirmed by the zero-field heat capacity measurement and the estimated Curie temperature is 7.6 K, which is lower than the magnetic transition temperature reported for the isostructural NdIr$_3$ ($T_C$ = 10.6 K)[22].

For PrIr$_3$, resistivity and heat capacity measurements reveal a clear anomaly at around 70 K. The same has been also reported by for NdIr$_3$[22]. By using the low temperature XRD structural analysis (between 20 K and 300 K), we conclude that this effect is related to a structural transition from the rhombohedral *R-3m* (#166) to the monoclinic *C2/m* (#12) structure. The structural transition revealed for PrIr$_3$, as well as NdIr$_3$, makes those two compounds different from their non-magnetic LaIr$_3$, CeIr$_3$ and ThIr$_3$ counterparts. Even more striking is the saturation of the volume change, which for PrIr$_3$ and NdIr$_3$ is not associated with the magnetic transition, although the magnetic character of Pr and Nd ions is likely to play an important role and should be clarified in future experimental and theoretical studies.

**Acknowledgments**

M.A.M. acknowledges support from the U.S. Department of Energy, Office of Science, Basic Energy Sciences, Materials Sciences and Engineering Division (low temperature x-ray diffraction analysis). TK thanks Piotr Twardowski and Mateusz Hajdel for helping to measure physical properties of PrIr$_3$ at Gdansk Tech. Research performed at Gdansk Tech. was supported by the National Science Center (Poland), project No. 2022/45/B/ST5/03916. Research performed at Princeton University was supported by the Gordon and Betty Moore foundation, grant number GBMF-9066.

# Supplementary Materials

**Ferromagnetism and structural phase transition in rhombohedral PrIr3**

Karolina Gornicka[1,2], Michal J. Winiarski[2], Robert J. Cava[3], Michael A. McGuire[1], and Tomasz Klimczuk[2]

[1]Materials Science and Technology Division, Oak Ridge National Laboratory, Oak Ridge, Tennessee, United States
[2]Faculty of Applied Physics and Mathematics and Advanced Materials Centre, Gdansk University of Technology, ul. Narutowicza 11/12, 80-233 Gdańsk, Poland
[3]Department of Chemistry, Princeton University, Princeton, New Jersey 08540, United States

SM. TABLE 1. The lattice parameters for PrIr$_3$ obtained in the temperature range 300 - 20 K.

| R-3m (#166) | | | | |
|---|---|---|---|---|
| Temperature (K) | $a_r$ (Å) | $c_r$ (Å) | | |
| 300 | 5.3344(2) | 26.268(1) | | |
| 250 | 5.3320(2) | 26.264(1) | | |
| 200 | 5.3295(2) | 26.261(1) | | |
| 150 | 5.3269(2) | 26.258(1) | | |
| 100 | 5.3246(2) | 26.256(1) | | |
| 90 | 5.3243(2) | 26.255(1) | | |
| 80 | 5.3240(2) | 26.254(1) | | |
| C2/m (#12) | | | | |
| Temperature (K) | $a_m$ (Å) | $b_m$ (Å) | $c_m$ (Å) | β(°) |
| 70 | 9.233(1) | 5.3159(9) | 9.265(1) | 109.17(1) |
| 60 | 9.2380(1) | 5.3132(8) | 9.260(1) | 109.10(1) |
| 50 | 9.2410(1) | 5.3116(7) | 9.259(1) | 109.08(1) |
| 40 | 9.2438(9) | 5.3095(5) | 9.258(8) | 109.07(1) |
| 30 | 9.2460(9) | 5.3089(5) | 9.2572(8) | 109.06(1) |
| 20 | 9.2477(9) | 5.3083(5) | 9.2566(8) | 109.04(6) |

SM. TABLE 2. The lattice parameters for NdIr$_3$ obtained in the temperature range 300 - 20 K.

| R-3m (#166) | | | | |
|---|---|---|---|---|
| Temperature (K) | $a_r$ (Å) | $c_r$ (Å) | | |
| 300 | 5.3243(3) | 26.237(3) | | |
| 250 | 5.3223(3) | 26.234(3) | | |
| 200 | 5.3199(3) | 26.231(3) | | |
| 150 | 5.317(3) | 26.227(6) | | |
| 100 | 5.3148(3) | 26.225(3) | | |
| 90 | 5.3144(3) | 26.224(6) | | |
| 80 | 5.3135(3) | 26.221(3) | | |
| C2/m (#12) | | | | |
| Temperature (K) | $a_m$ (Å) | $b_m$ (Å) | $c_m$ (Å) | β(°) |
| 70 | 9.2234(7) | 5.3052(6) | 9.2480(5) | 109.085(2) |
| 60 | 9.2252(7) | 5.3038(5) | 9.2462(4) | 109.059(2) |
| 50 | 9.2285(6) | 5.3018(5) | 9.2458(5) | 109.07(1) |
| 40 | 9.2291(7) | 5.3009(5) | 9.2455(5) | 109.07(1) |
| 30 | 9.2301(7) | 5.3005(5) | 9.2449(6) | 109.06(3) |
| 20 | 9.2311(7) | 5.3007(4) | 9.2446(6) | 109.05(6) |